**Fireball streak detection with minimal CPU processing requirements and the automation of the Desert Fireball Network data processing pipeline.**

Martin C. Towner[1]*, Martin Cupak[1], Jean Deshayes[2], Robert M. Howie[2], Ben Hartig[2], Jonathan Paxman[2], Eleanor K. Sansom[1], Hadrien A. R. Devillepoix[1], Trent Jansen-Sturgeon[1], Philip A. Bland[1]

[1]Department of Applied Geology, Curtin University, GPO Box U1987, Perth Western Australia, 6845, Australia
[2]Department of Mechanical Engineering, Curtin University, GPO Box U1987, Perth Western Australia, 6845, Australia
*corresponding author's email: martin.towner@curtin.edu.au

# Abstract

The Desert Fireball Network (DFN) uses a network of cameras to track and triangulate incoming fireballs to recover meteorites with orbits.< the full piple ine form detection ot tringuaiton is automated to reduce human requirtemetnrts to run a large netowkr. Specifically the first stemp in automatinis the detection algoritm.>

The detection of fireballs streaks in astronomical imagery can be carried out by a variety of methods. Fireball detection is done on-camera, but due to the design constraints imposed by remote deployment, the cameras are limited in processing power and time. We describe the processing software used for fireball detection under these constrained circumstances. Two different approaches were compared: 1) A single layer neural network with 5 hidden units that was trained using manually selected fireballs, and 2) a more traditional computational approach based on cascading steps of increasing complexity, whereby computationally simple filters are used to discard uninteresting portions of the images, allowing for more computationally expensive analysis of the remainder. Both approaches allowed a full night's worth of data (over a thousand 36-megapixel images) to be processed each day using a low power single board computer. We distinguish between large (likely meteorite dropping) fireballs, and smaller fainter ones (typical 'shooting stars'). Traditional processing and neural network algorithms both performed well on large fireballs within an approx. 30000-image dataset, with a true positive detection rate of 96% and 100% respectively, but the neural network was significantly more successful at smaller fireballs, with rates of 67% and 82% respectively. However, this improved success came at a cost of significantly more false positives for the NN results. Simple consideration of the network geometry indicates that overall detection rate for triangulated large fireballs is estimated to be better than 99.7% and 99.9%, by ensuring that there are multiple double-station opportunities to detect any one fireball.

<after detciton, data is passed to a central server, where correlations betqeen cameras are used to filter fals events, and … >

# Keywords

Hough transform, image processing, astronomy

# Introduction

## Background and overview

Fireball camera networks consist of a widely spaced array of astronomical optical instruments designed to observe incoming meteors leading to the recovery of fresh meteorite falls, with an associated orbits. A fresh, uncontaminated meteorite associated with a known orbit provides valuable information about the composition and formation of the Solar System.

Several major fireball networks have been in operation throughout recent history, and today many regional networks exist utilising still and video technology, including the Desert Fireball Network (DFN) in the Australian outback (Bland et al. 2012; Weryk et al. 2007; Oberst et al. 1998; Trigo-Rodríguez et al. 2006; Colas et al. 2014; Jenniskens et al. 2011; Cooke and Moser 2011). The DFN has unique constraints and advantages over other networks; the Australian continent is ideally suited to meteorite recovery, due to large areas of low vegetation with pale colour rocks, making meteorite finding easier, and good weather, giving both a greater percentage of clear nights for observations, and also lower chance of meteorite contamination due to precipitation. Conversely the large distances and remote nature of the Australian outback pose logistical and communication challenges.

Considering data analysis, early fireball networks (the earliest of which began in the 1960s) were film based, using long exposures, typically all night. Images were inspected manually, and negatives with fireballs were then scanned and processed by hand (Spurný, Borovička, and Shrbený 2006). For film systems, a supplementary brightness sensor, such as a photomultiplier tube, is used to detect events and provide time brackets for an operator to review the films. Images were then calibrated by comparison of the streaks left by long exposure stars with the known star positions and exposure start/end times. Data reduction and fireball detection involved considerable manual activity for each fireball.

More recently in the digital age, video systems are prevalent, and there are several extant software tools specifically for fireball detection, such as METREC (Molau 1998) or more recently UFOCapture (Sonotaco, Japan) or ASGARD (Brown et al. 2010). One can also use generic motion detection software, as used on many security camera systems. Video systems have sufficient resolution to usefully calculate fireball orbits, but generally lack the resolution to capture fully the final stages of ablation. This precision is needed for meteorite recovery, unless the cameras are placed relatively closely together (typically less than 50 km), or the fireball is particularly big. This is impractical for the DFN, which covers large areas of the sparsely inhabited Australian outback. Instead of video, the DFN instead uses high resolution still images, from a commercial DSLR system (currently Nikon D810 taking 25 second exposures), and uses an innovative technique using a liquid crystal shutter to embed absolute timing information of the fireball within the 25 second exposure (Howie et al. 2015). The higher resolution of the DSLR compared to video allows a larger camera spacing of about 100-150 km, while still observing fireballs with enough accuracy to give a reasonable chance of meteorite recovery. Such a system still requires software to automatically detect fireball events, unless one is manually surveying many thousands of images. The core of event detection for video systems is around changing pixels, and motion detection. This can be coupled with real time smoothing/noise removal, and tracking of motion over multiple frames, to remove short-lived false positives. Real time processing of video systems requires significant processing and storage capability; for example, recommended systems requirements for UFOCapture are 2.4GHz Pentium 4 for a 640 x 480 resolution video system. In the case of long exposure still images, the 'number of frames' is significantly fewer, so more time can be devoted to analysing each image, but the basic approach used is similar: one searches for changes between frames, and localised and categorises them.

Machine learning was defined by Arthur Samuel in 1959 as a "field of study that gives computers the ability to learn without being explicitly programmed" (Samuel 1959). It has been implemented in a wide variety of application areas, from medicine to pedestrian detection including advertising (Perlich et al. 2014), finance (Clémençon, n.d.), military (DesJardins 1997), and astronomy (Ball and Brunner 2010). The potential of machine learning is being developed at every level within the era of big data management. Machine learning algorithms are categorised in three main classes: Supervised learning, unsupervised learning and reinforcement learning algorithms and for fireball detection a supervised learning approach appears best suited, as the algorithm can be presented with training data that has been categorised by a human. Additionally it is also one of the least computationally expensive algorithms, which matches well with the hardware requirements of solar-powered remote camera systems that make up the majority of the DFN observatories. In supervised learning, the computer is presented with a set of sample data and its corresponding output, forming a training set. The computer will then improve its prediction accuracy by reducing the error between the predicted value and the correctly labelled value, to reach a general model that best maps inputs to outputs (Bishop, 2006). In this case, a binary classifier: to indicate fireball or not-fireball when presented by an image.

There appears to be little previous application to fireball image detection to date. The only publication the authors are aware of is (Zhao 2010), who applied Support Vector Machine (SVM) classification algorithms to give the computer the ability to learn how to cluster fireball event data, obtained from radar observations. This work appears to have been successful in the identification of meteors with a trajectory perpendicular to the radar beam but had difficulties in the detection of other meteor trails due to the poor quality of the source data.

Machine learning has potential to improve data throughput, and better handle 'strange looking fireballs', where the possible variety of shapes, colours and textures of fireballs can cause difficulties for traditional processing methods that expect a simple straight-line, non-fragmenting fireball. However, when considering the trade-off between traditional approaches and machine learning, one must be cognisant that machine learning can be hardware intensive, which can be an issue for remote camera systems such as the DFNs', which rely on solar panels and batteries.

## Hardware constraints on remote camera systems

For the DFN systems, the cameras are located at remote sites across the Australian outback, with limited power and communication. The camera systems are typically powered by solar panels and batteries, and communications are via mobile data service, which is of somewhat erratic performance in central Australia. Hence, the deployed systems must be highly autonomous and robust, and operate with a low bandwidth whilst still providing timely information concerning any possible fireballs. The success of the project depends on coverage of a large area at optimal cost, which requires limited team members and many cameras, leading to the driving requirement of a highly automated, low cost camera network.

Due to bandwidth constraints, the full resolution images cannot be transferred and processed centrally. This leads to the requirement of being able to carry out on-board processing of a full night's worth of images every 24 hours, to keep up with data collection. Instead, only low volume event data such as lists of times and coordinates can be transferred. In opposition to this requirement for processing high volumes of data, the remote deployment requires that the system is relatively low power, to reduce unit cost of ancillary power systems. This leads to use of a low power, low cost, single board computer (currently Commell LE-37G using 15W). In turn, this low processing capability and time constrains lead to the development of energy-efficient techniques of fireball detection.

# Methods

## Traditional computation

Here we describe an event detection method based on a chain of image processing operations. The philosophy that has evolved is to process each still image using a chain of increasingly computationally expensive operations on fewer and fewer pixels, such that simple initial tests are used to exclude as much of the image as possible early in processing. The software is implemented in Python, using OpenCV, SciPy/NumPy and the scikit-image libraries (Oliphant 2007; Jones,

Oliphant, and Peterson 2001; van der Walt et al. 2014). The use of Python with NumPy allows C comparable performance but reduced development time, and ability to run on a variety of platforms. A variety of chains of operators were tested, and timed using test data sets. To optimise the CPU usage for the system, the image under analysis is split into 400 x 400 pixel tiles, which can each be handled in parallel by CPU cores. Each tile is processed independently by comparison with the corresponding tile from the previous image. The fisheye lens used by the DFN systems means that the astronomical image does not completely fill the camera sensor field of view (Figure 1), and hence as a first step, blank tiles outside the astronomical image can be immediately discarded. Masking of the image is also applied, to discard parts of the image where the view of the sky is obstructed. By processing the image as separate tiles, one benefit is that it is easier to filter and handle varying background brightness across the whole image, as low frequency spatial variations can be treated as constant across one tile. This is a classic issue with fisheye all sky images, and is particularly relevant near to sunrise/sunset, where sky brightness is varying greatly across the image.

Figure 1 – Camera image showing a

Figure 2 – DFN data pipeline

## Details of the event detection procedure

The system takes still images, and the fireballs we wish to detect within them are transient phenomena. The initial operation in the chain of elimination is to calculate the difference between the current and previous image tiles, and quantify any changes seen.

Step (1):

We begin by calculating a brightness difference between each pixel in the tile and its corresponding pixel in the previous image's tile, and counting the number pixels that have a difference above a threshold value. The use of a threshold of difference allows small variations in brightness due to slight atmospheric or sensor issues to be accounted for. Counting numbers of anomalous pixels rather than summing their differences prevents any one anomalous pixel change to dominate the result. If no or few pixels are changed by less than a given amount, then the tile can be discarded immediately. Tiles containing the moon are essentially saturated, due to the high camera sensitivity settings, and as such will immediately fail the first difference operation, and not be passed to subsequent operations.

Step (2):

This differencing operation is then repeated, but with a Gaussian blur applied independently to the current and previous tile prior to differencing. This blur operation smears objects that have moved slightly, such as stars far from the celestial pole, reducing the differenced values around such objects below the detection threshold. Tiles that drop below this blurred threshold are then discarded.

Step (3):

The remaining few tiles with significant differences between current and previous are then analysed using a Hough Transform (Duda and Hart 1972) to detect straight lines in the tile and extract the pixel coordinates of any line.

This involves pre-processing and thresholding according to the following processes:

(i) Absolute difference between current and previous tiles.

(ii) The Hough transform requires a binary image, so an Otsu threshold (Otsu 1979) is used to convert the greyscale difference to binary. The Otsu algorithm chooses a threshold based on attempting to fit two Gaussian brightness populations to the brightness histogram of the image tile. The Otsu thresholding is well suited to images containing star fields and fireballs, which essentially consist of bright points (and a bright fireball) and a dark background.

(iii) Applying a Skeletonize function to reduce blurry, smeared lines (due to lens defects) down to single pixel lines, giving the Hough transform sharper edged lines to work with. Short false lines such as smeared stars are reduced to single points, and not detected by the subsequent Hough transform.

(iv) A 3-pixel dilate is then applied to the skeletonized line, to thicken the lines, such that the Hough transform will easily detect lines without having to finely tune the Hough voting parameters.

The Hough transform function can then be applied at a relatively high sensitivity setting, which becomes more computationally expensive as more pixels have to be checked for each putative line. But this is acceptable as only a few tiles are to be analysed. Hough parameters are chosen based on the ability of the transform to detect both dashes and long lines.(In the DFN systems, the cycling of a liquid crystal shutter is used to break the fireball track into countable dashes to provide velocity data (Howie et al. 2015)). The output of the Hough transform is a list of (start,end) pixel coordinate pairs for each line found within a tile.

Step (5):

Following this line detection, a series of ad-hoc rules are applied to reduce the false positives within the lists of lines detected. These rules are based around the physical expectations of likely fireball events. By applying a series of simple rules, one after the other, many false positives are removed at little computational expense:

i) Pixel brightness is considered along the length of the line; it should be relatively constant, or peaking at the line centre. A short line with high variance, with brightness peaks at line ends, will correspond to a false line fitting through two very close bright stars, and is discarded. However, care must be taken in the choice of variance thresholding, to ensure that dashed lines are not discarded. Additionally, within a tile, a search is carried out for lines that are very similar (similar start and points) to remove double counting and duplicates.

ii) The line coordinates from all tiles in one image are then collated together, and filtered based on plausibility checking of the resultant list. If there is just one very short line in the whole image, e.g. five pixels long or less, it is discarded as a false positive. (In reality, this could represent a fireball, but of too small or distant a result to be useful for meteorite recovery or orbit determination). Stationary or slow moving objects are also removed, based on coordinate listings from earlier images in the sequence. Objects that appear in the same spot or moving slowly and linearly (e.g. a plane, or the International Space Station) over three or more images have persisted for too long to be a fireball, and are discarded.

iii) Finally, (at the single camera level) ad-hoc rules based on experience with specific problem are applied: In some cases the Hough transforms returns large numbers of lines for a single image (for example on a partially cloudy night, detecting cloud edges); hence, results are discarded if over 10,000 lines are observed in one tile. This has had the side effect of sometimes discarding the oversaturated middle sections of particularly large fireballs, but the fainter beginning and end of a fireball are still correctly detected (which is acceptable from a fireball detection and triangulation emphasis). Other issues are caused by cloud edges when they move between subsequent images as they are detected by the Hough transform as linear features. Many can be removed, as they are often persistent over multiple images and the direction of the line detected is not the same as the direction of the apparent motion of the line between images, as would be the case for a fireball.

This process is intended to efficiently detect as many real events as possible, and not miss any reasonable candidates. As such, it still generates significant numbers of false positives. From the DFN fireball point of view, this is acceptable, as it is better to have 10 false rather than one important event missed.

## Testing dataset

For validating the algorithms and selecting optimum parameters, we have assembled a selection of 50 image sets to cover most plausible scenarios. These included genuine fireballs and likely false positives of a variety of sizes such as planes and satellite passes, over a variety of cloudy and clear nights, plus nights with no fireballs present (moonless and with full moon) and situations such as a bright star rising/setting, or passing behind an obscuring tree (and hence flickering between images). Examples of some of these are shown in Figure 3.

The Hough transform is designed for finding linear features, such as fireballs, but the parameters must be chosen carefully to optimise for the character of the lines desired (typically length, thickness, 'straightness', permitted length of breaks in line), and this dataset allowed us first to test procedural changes, such as adding more filtering steps, and secondly to refine the chosen parameters.

The parameters used in the algorithms were adjusted until all features were correctly identified (or discarded) in the test dataset.

Figure 3 – Test dataset images, showing (a) plane streak, (b) satellite streak, (c) straight cloud edge and Moon aperture diffraction spikes, (d) small fireball

Table 1 - list of Parameters

# Neural Network based detection method

## Preprocessing

The dimensions of the images captured by the DFN cameras are 7360x4912 pixels, which corresponds to 36,152,320 pixels for each colour channel. Any machine learning algorithm would struggle with a vector of this size, so the image was initially split to extract only the green channel (as is also done for the traditional processing). As in the traditional processing, to make the problem mo9re manageable, the image is then split into tiles, and those tiles are classified. A tile size of 25 x 25 pixels gives a state vector of length 625, which is suitable for a simple neural network classifier. However, a tile of this size is a very small part of the sky, typically less than one degree of viewing angle. Hence to ensure that 25 x 25 is a large enough area of the image to encompase reasonable freatures—such as fireball streak—before tile generation, the image was downsampled by factor 2 using bilinear interpolation. Hence the tile now represents approximately 2 x 2 degrees, albeit at a lower resolution. Given the software goal is detection of fireball in a tile, not precise position of that fireball, and meteorite dropping fireballs will not be faint, this compromise appears reasonable.***

For the NN classifier, as in the trad processing, pre-processing significantly improves the true positive rate. In particular for NN, reducing the image background noise will reduce spuriios detections. Thresholding was done in a slightly different manner to the traditional approach above; the previous image was blurred, and ssubtracted from the currunt image, and then reciprocally the current image was blurred, and subtracted from the previous. binary dilation was appled independently to each image of the new pair, before the two images were compared. This process effectively allows for changes in background brightness between images (such as closer to sunrise/set), and also blurs out and removes star edges (which may have shifted slightly between images), preventing them from generating false postivies, leaving only major differences.

Another benefit of this approach to image subtraction and filtering was to address a few issues encountered by the traditional detection code. It was noticed that features such as three close stars linearly arranged would often produce a false positive identification, interpreted as three dashes of a fast meteor streak. This was particularly an issue with stars far from the celestial pole, such that there was significant movement between frames, so that basic image differencing was imperfectly removing them leaving residual peaks that were misinterpreted.

## Neural Network details

By breaking the original image into 25 x 25 pixel tile, this generates an input vector to the neural network of 625 integers. The network requires one output unit, giving a 0 or 1 for a tile, of 'fireball' vs 'not-fireball'.

It was decided to code neural network in Python, rather than using one of the existing implementations, to give more control over minimizing resource use. A single layer, forward propagating neural network was constructed with a gradient descent optimisation minimize function (from SciPy (Jones, Oliphant, and Peterson 2001)), with back propagation explicitly implemented in training.

The choice of the optimal number of hidden layers and nodes in a neural network for a particular task is currently not an analytically solved problem, and is usually approached heuristically. Conventionally, the number of hidden nodes should be between the number of output and input units—so from 1 to 625—and ideally it is dependent on the number of training examples <ref>. We have carried out an initial investigation and testing using a training dataset (detailed below) to test performance versus computation time over the range of 1- 10 hidden nodes. Under 5 hidden nodes the network was suffering from high bias, as the model could never reach a higher accuracy than 70% on the training set. Above 5 hidden nodes, the difference in the outcome was so small that it was difficult to choose the best parameter. Above 10 hidden nodes issues arose due to excessive time and computational requirements. Hence, 10 hidden nodes seemed reasonable to achieve high performance without considerably suffering from high variance.

***<?The regularisation term in the modified cost function provides a way to penalise the cost function in order to control overfitting. In this case, trials with comparing the cost of fitting the training and cross-validation datasets show close convergence for regularisation parameters above $2^8$, so this was chosen as the value used for network training.>

Within the overarching pipeline—similar to the traditional processing pipeline—ad hoc post processing operations helped to improve the algorithm success rate. The tiles were classified by the network into fireball or not-fireball, and results for each image were collated, and coordinates processed in identical manner to the Hough transform-based detection, such that slow moving objects seen in three consecutive images were ignored. Coordinates for the fireballs seen are represented as centre of fireball-classified tiles, and so are accurate to 25 pixels. This is not as precise as the Hough technique (resulting in typically 5-10° of altitude/azimuth pointing accuracy) but good enough for central server to carry out preliminary triangulations to check for the validity of proposed trajectory.

## Training and test dataset

Successful learning of a neural network is achieved by the quality and quantity of the materials provided to the network along with the right parameter selection. When dealing with numerous features, the datasets are usually partitioned in three parts: the training, cross validation and test dataset. These sets are strictly independent from each other but made of similar examples that could have been used for any set. The three datasets were initially prepared by collecting some examples from manually searching the images. About 50 images of manually selected fireball events were pre-filtered giving 200 tiles containing fireball streaks. The same quantity of negative examples were obtained from noisy images containing background noise or unidentifiable parts of meteor streaks as shown by the images below (Fig <x>). Training based on this dataset showed that the cost function not fully stabilised. Hence the training set was expanded by factor of eight, by copying, flipping and/or rotating each example tile. The datasets were then made of an equal proportion of positive and negative samples randomly distributed to each dataset with 60% to the training set, 20% to the cross validation set and 20% to the test set.

Figure 4—Subset of positive and negative samples for neural network training

<training results>

# Testing results and discussion

<Initial evaluation was to test the performance of any algorithm operating on the correct hardware as deployed in the DFN cameras. The algorithm must ba avle to process a full nights worth of images during the day, to keep up with the data volumes generated. Initial testing showed this was not a major concern for all plausible variations on the algorithms. Typically the average processing time for an image was 6sec for hough and 5 for nn, resuling in processing a full nights data in about 2 hours.>

To test the performance of the on-board camera detection procedures, a month of real observations from one fireball camera (24457 images) were manually surveyed for fireballs and satellite events, and these observations were then compared to results using the traditional and neural network event detection procedures.

Details of the manual and automated testing results for the month's data are shown in Table 2. The data appears as seven rows showing groupings of five nights of observations, to indicate the variability seen. Summary statistics are shown at the base of the table. Some five night groups, such as the 11-15 block—corresponding to November $15^{th}$ to $19^{th}$, 2014—have relatively few images present, showing several fully or partially cloudy nights of observations. During cloudy nights, the camera system takes fewer images, to minimise unnecessary data storage. In both the manual surveying and software-based detection, satellites are counted as valid events equivalent to a fireball, and no distinction is made: Satellite streaks are valid events from the point of view of a single camera, and detection and triangulation of satellites remains a relevant scientific interest for the camera network, in addition to the primary goal of meteorite recovery. From the summary, for all fireballs the algorithmic detections have a success rate of about 70%, compared to manual surveying. This at first appears less than ideal, given the high priority of avoiding true negatives at the expense of false positives. However, the manual surveying is recording all fireballs, from the very smallest seen by the eye upwards (see for example Figure 5(a)). Given that the overarching rationale for the fireball detection is to recover meteorites, the software's 'goal' is to highlight large fireballs. Therefore, to investigate this further in the data table we also highlight large fireballs only. Large is defined as appearing on two or more adjacent tiles, or having a brightness that is saturating the sensor at the brightest spot; for example Figure 5(b,c). In this case, the automated true positive rate is about 96% Hough/100% Neural, more in line with the project requirements. Of the 51 large fireballs observed, the traditional software missed two: they were both long streaks from satellite traces, where the brightness of the trace was relatively low which resulted in a failure to highlight the streak at the Otsu thresholding stage (Figure 5(d)). The Neural Network version detected all large fireballs.

The 5 night group 11-25 to 11-30 has high number of false positives (278 for the Hough algorithm, 1560 for the neural network). On the night of 11-29, there was a thunderstorm on the horizon, which resulted in lightning repeatedly illuminating the edges of distant clouds producing transient bright lines on the image, which was erroneously flagged as events by the software (Figure 5(e)), as they satisfy the criteria of a single 'straight' line, with constant brightness along the line, and nothing in the prior and subsequent images. This produced over 70 false positives in one 3 hour period. Fortunately, this combination of straight cloud edge near the horizon with internal lightning appears relatively infrequently, as closer storms would not give the same appearance. As described in the following section, such events are filtered from the data pipeline at the camera-to-camera triangulation stage of analysis.

False positives from the on-camera detection algorithm appear overall at the rate of approximately 4 times the real events for the Hough algorithm, but approximately 10 times for the NN. <comment on training completnress>. Other examples of false positives are beams of light from camera internal reflections, and cloud edges as previously discussed. Almost all of these false positives are discarded at the camera-to-camera level, as they fail to match with a corresponding event on an adjacent system. Satellite observations as opposed to genuine fireballs are of interest for a variety of studies. They can be separated from fireball event alerts from calculation at the multiple camera level, whereby triangulation of the data will indicate the altitude and orbit of the object. (Also allowing removal of low level objects such as planes).

Table 2 – Analysis of 5 weeks of imagery from a single camera, in blocks of 5 days

Figure 5 – results

# Event detection at the multiple camera level within the fireball network.

The automation of the full chain of data processing from single camera event detection to server level outputs is an essential component of a large camera network such as the DFN, due to the large data volumes. Following the single camera event detection, the events seen on the system are converted from (pixel x, y, time) coordinates to astronomical coordinates using a predetermined calibration function, and then transferred as simple text files to the DFN central server, which then collates events, looking for multiple camera observations of each possible fireball.

Several known false positive scenarios will be passed by the single camera event detection, and successfully removed from the putative events at the multiple camera level. Imagery of aeroplanes usually lasts more than 30 sec, stretching over three images, and are thus easy to filter. This will not discard a small satellite flare or plane that lasts less than 30 sec. If the event is seen by an neighbouring camera system, triangulation of the feature gives an altitude that is too high, in the case of satellites, and too low in the case of planes (Although planes are so low that they are unlikely to be seen on multiple cameras simultaneously).

Following triangulation as part of main data processing pipeline, preview images and coordinate data for any successful triangulations are sent by email for manual confirmation and data point extraction. After validation (or removal of false positives), these remaining events are stored on the central server, and software automatically collects and calibrates nearby related imagery for further analysis. One further human interaction step remaining is the precise marking of dashes for the fireball track to accurately timestamp each dash. After this, the rest of pipeline is automated, producing final orbit results, and if reasonable, predicted fall positions for all events, whilst updating network status records, such that team members can easily concentrate on more in-depth processing of the most promising meteorite-dropping candidates.

## Overall detection rates

For the detection of a fireball in a two camera network, each detecting independently, the probability of detection is P(A) x P(B), where P(X) is probability of fireball detected at camera X. In the simplest case P(A) = P(B), and for large fireballs P(A) is 0.958 (Hough) or 1.0 (NN) from Table 3, giving two camera fireball detection as $P(A)^2 \approx 0.92$ or 1.0.

However, the reality of more cameras means that fireballs are rarely (only in literal network edge cases) going to appear on only two cameras, and viewing distance and conditions will break the equality. We can illustrate this with consideration of two examples: In the case of 4 cameras in a square, with a fireball in the middle, the probability to successfully detect a fireball pair becomes one minus the probability of all cameras failing to detect the event, plus the possibility of 4 combinations of 3 cameras failing to detect the event (such any two cameras detecting the event will count as successful result). Where the failure of a camera to detect an event is 1-P(A), so this calculates explicitly as:

$$1 - [(1 - P(A))^4 + 4(1 - P(A))^3] \tag{1}$$

Which evaluates to 0.9997 (Hough) and 1.0 (NN) with the values above.

In a second illustrative examples, we can consider the hypothetical case of a triangular pattern camera network (Figure 6), the two 'primary' cameras (A) close to fireball give a probability of P(A) each, and surrounding secondary cameras (a) would have a lower detection efficiency due to the greater distance of the fireball event (denoted P(a)). A successful triangulation occurs when any two cameras detect the event. So the detection success of the array can be estimated as a one minus the combinations of failures-to-detect from the 2 primary cameras and 8 secondary cameras. For 2 primary and N secondary, we write P(2Pf, N-1) as the probability of both primaries failing and N-1 secondaries failing, and similarly P(1Pf, N) is the probability of 1 primary failing and all secondaries failing. These can be evaluated as:

$$P(2Pf, N - 1) = (1 - P(A))^2 . N(1 - P(a))^{N-1} \tag{2}$$
$$P(1Pf, N) = (1 - P(A)) . (1 - P(a))^N$$

Giving:

$$P(\text{successful triangulation}) = 1 - [P(2Pf, N - 1) + P(1Pf, N)] \tag{3}$$

Using the small fireball detection estimate from Table 3 of 0.66 (Hough) and 0.82 (NN), we can estimate (3) as 0.997 (Hough) and 0.99995 (NN).

In this network configuration, there is also the network edge case, where the fireball is outside the centre of the network, and seen close to two cameras and more distantly by the surrounding 5 cameras. In this case (3) gives the probability of success as 0.94 (Hough) and 0.995 (NN).

In these simple illustrative calculations, we can see that despite a lower small fireball detection threshold for individual systems, the advantage of many combinations of cameras results in a high triangulation reliability. Note this is only the detection rate when the cameras are operating: To calculate the full true detection rate of the network in order to derive an Earth fireball flux, consideration must be made of cloud coverage, twilight, moon saturation, operational failures, brightness, and distance from observatories, as discussed in a future paper in preparation.

# Conclusions

The DFN consists of an array of remote astronomy camera systems designed to observe and detect incoming fireballs, with the goal of recovering meteorites. The observatory cameras are constrained by power and cost, in order to maximise ground coverage and hence meteorite recovery chances. Since the cameras are operating remotely and independently, the images collected during a night must be processed on-board for the detection of fireball trails. We have developed an image processing chain optimised for small power systems, whereby a cascading series of more computationally expensive operations are applied to smaller and smaller regions of interest. We implement this as a tool chain of initially simple operations like subtraction or Gaussian blur, followed by a Hough transform. Supplementary ad-hoc rules to remove objects, like slow moving planes, reduce the number of false positives. In parallel, for comparison we have also implemented a simple neural network approach, based on a single layer forward propagator with hidden 10 nodes, again with pre-processing to improve success rates, and ad-hoc post processing to remove some false positives.

Testing of both of these approachs using a trial dataset of raw data collected over one month of operations indicates that this is an efficient method for use in situations of constrained computing, such as low power equipment. The traditional method was slightly less effective than the neural network, although the NN generates significantly more false positives without sacrificing true positives, giving a large fireball, single camera event detection rate of 96% compared to manual observations of data. <the nn has been fully trained, according to xxx test>. Both the traditional and neural netowkr processing are most effective with larger fireballs (defined as appearing on multiple tiles in the image detection algorithm), but have a lower success rate with smaller, fainter meteors (62%). However, the topology of the camera network means that several cameras have the opportunity to observe an event, so even a low success rate for single camera-small fireball cases still results ina high netowkr fireball detection rate; such that the critical metric for a meteorite recover network  is that the detection rate is better than 99.9% for large fireballs***.

The primary goal of the DFN project is the recovery of meteorites; hence processing has focused on correctly identifying all large fireballs capable of dropping meteorites, at the expense of generating false positive events. The majority of these false positives can be discounted at a later level of processing, such as attempting to triangulate contemporaneous events from multiple cameras; in the current operations, human interaction is needed for about 20 events per day (most of which are false alarms), from a camera network of 50 cameras, each producing about 10-20 false and typicall 1-2 real events per night.

# Tables

*Table 1 - list of Parameters selected for the event detection software, as detailed in the processing steps.*

| | |
|---|---|
| Tile differencing settings | |
| Tile size | 400 |
| Minimum Threshold between two pixels when difference tiles | 50 |
| No of pixels required to exceed threshold in a tile for tile to be considered | 100 |
| # same thing but after applying median blur | |
| No of pixels required to exceed threshold in a tile for tile to be considered after Gaussian blur and difference | 100 |
| Gaussian blur pixel size chosen | 5 |
| Hough settings | |
| Distance resolution of the accumulator in pixels ('rho') | 1 |
| Angle resolution of the accumulator in degrees | 1 |
| Accumulator threshold parameter. Only those lines are returned that get votes > threshold. | 5 |
| Minimum line length. Line segments shorter than that are rejected. | 14 |
| Maximum allowed gap between points on the same line to link them. | 9 |
| Ad hoc and line filtering settings | |
| Max acceptable standard deviation in brightness along the line, compared to average. | 1 |
| Minimum average brightness along a line. | 11 |
| No of lines required in a tile for tile to discarded as not fireball | 20000 |
| Isolated single line minimum length (drop any imgs with isolated line which is only this long) | 35 |
| Pixel distance a line must be from others to be 'isolated' | 500 |
| Pixel separation between two lines, start-start and end-end for line to be considered 'similar' | 20 |
| Max permitted distance between start and end of lines in consecutive images to be considered continuation of the same line (for detecting for example planes) | 30 |
| Neural Network values chosen | |

Nodes

<Training etc>

*Table 2 – Analysis of 5 weeks of imagery from a single camera, in blocks of 5 days, from late 2015.*

| MM-DD | Number of images. | Fireballs manually observed. | Hough detections (inc. false positives). | Genuine Hough detections. | Neural detection (inc. false postives) | Genuine Neural detection | Hough Detection rate, all (%) | Neural Detection rate, all (%) |
|---|---|---|---|---|---|---|---|---|
| 10-25 | 7628 | 56 | 107 | 40 | 1447 | 49 | 71.4 | 87.5 |
| 11-00 | 2837 | 25 | 51 | 19 | 133 | 21 | 76.0 | 84.0 |
| 11-05 | 2579 | 33 | 69 | 33 | 471 | 35 | 89.2 | 94.6 |
| 11-10 | 3954 | 20 | 29 | 11 | 475 | 14 | 50.0 | 77.8 |
| 11-15 | 126 | 1 | 1 | 0 | 52 | 0 | 0 | 0 |
| 11-20 | 3555 | 29 | 82 | 20 | 1563 | 26 | 69.0 | 89.7 |
| 11-25 | 3778 | 35 | 278 | 29 | 1560 | 31 | 76.3 | 81.6 |
| | | | | | | | | |
| Totals | 24457 | 199 | 617 | 150 | 5701 | 176 | 73.5 | 86.3 |

*Table 3 – Data subdivide into long/big fireballs and remainder ('small fireballs'). Long/big fireballs are defined as greater than one Hough size image tile (400 x 400 pixels).*

| MM-DD | Long fireballs (manually observed). | Genuine Hough detections, long. | Genuine neural detection, long | Hough detection rate, long, (%) | Hough detection rate, small fireballs only, (%) | Neural detection rate, long, (%) | Neural detection rate, small, (%) |
|---|---|---|---|---|---|---|---|
| 10-25 | 7 | 7 | 7 | 100.0 | 67.3 | 100.0 | 85.7 |
| 11-00 | 11 | 11 | 11 | 100.0 | 57.1 | 100.0 | 71.4 |
| 11-05 | 13 | 12 | 13 | 92.3 | 87.5 | 100.0 | 91.7 |
| 11-10 | 3 | 3 | 3 | 100.0 | 40.0 | 100.0 | 73.3 |
| 11-15 | 0 | 0 | 0 | NA | 0 | NA | 0.0 |
| 11-20 | 5 | 4 | 5 | 80.0 | 66.7 | 100.0 | 87.5 |
| 11-25 | 9 | 9 | 9 | 100.0 | 68.9 | 100.0 | 75.9 |
| | | | | | | | |
| Totals | 48 | 46 | 48 | 95.8 | 66.6 | 100.0 | 82.1 |

# Figure captions

*Figure 1 – Camera image showing a 3 second fireball seen relatively closely to the camera, at Perenjori in Western Australia, on 27$^{th}$ April 2015. Such an event is easy to detect, although extra false positive coordinates were produced due to the presence of moon and moon halo close to track.*

*Figure 2 – DFN data pipeline showing processing steps and decisions for both the traditional Hough transform based processing and the Neural Network processing.\*\*\**

*Figure 3 – Test dataset images, showing (a) plane streak, (b) satellite streak, (c) straight cloud edge and Moon aperture diffraction spikes, (d) small fireball*

*Figure 4—Subset of positive and negative samples for neural network training*

*Figure 5 – results of auto detection (Hough and NN), showing examples of a variety of small fireballs and false alarms. (a) small fireball not observed by auto-detection, (b), (c) fireballs detected by auto-detection, (d) one of the two 'large fireballs' not seen by Hough-based detection algorithm (actually a relatively dim satellite streak), (e) lightning-illuminated cloud on the horizon that produced a false positive.*

*Figure 6 – Idealised ground layout of camera network in a triangular pattern, with 2 central primary cameras (solid circle) and surrounding secondary cameras (open circle).*

# 13 Figures

14 Figure 1

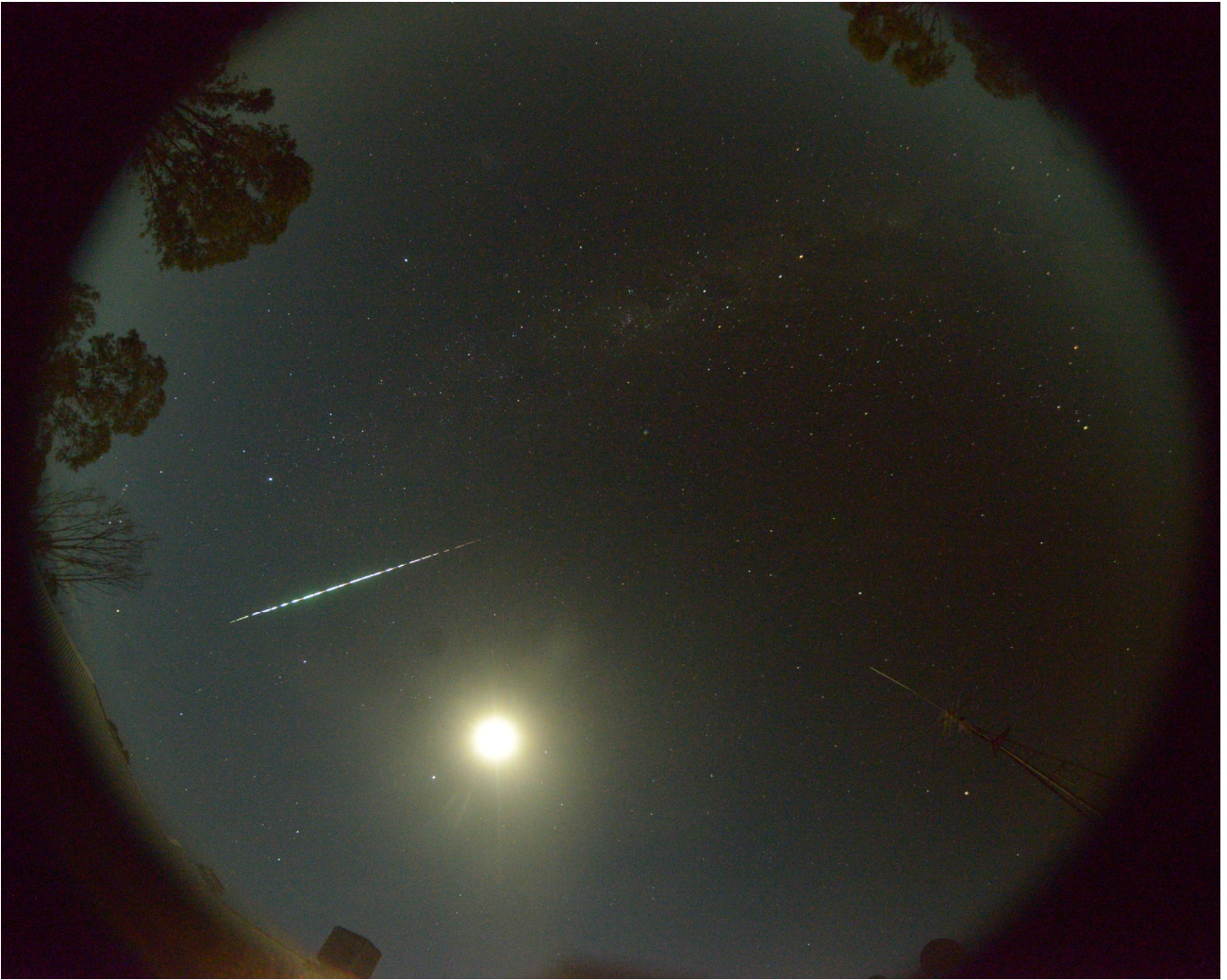

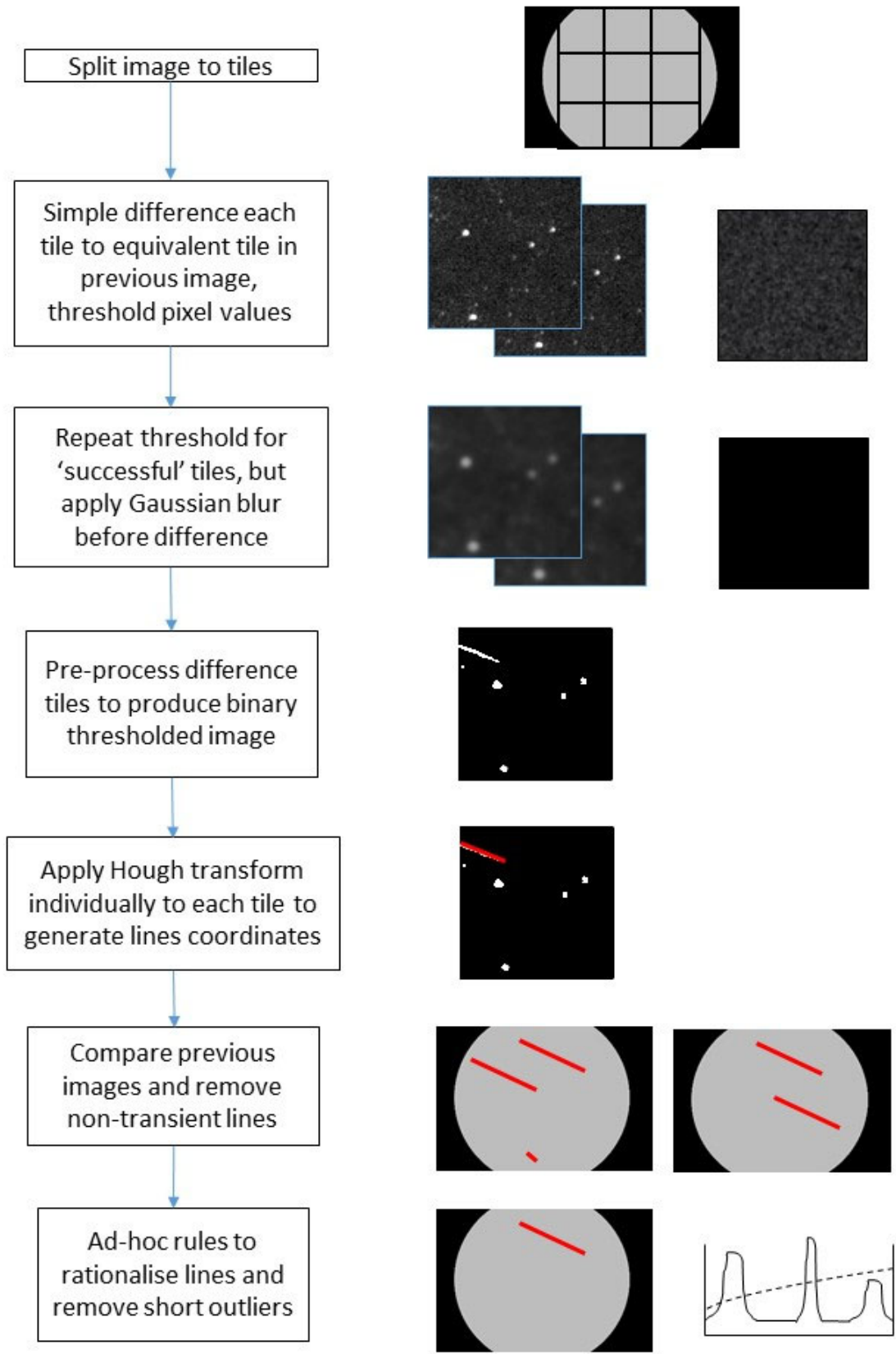

21   Figure 3

(a)

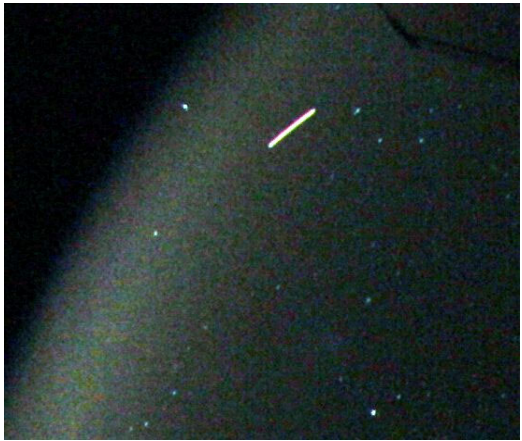

(b)

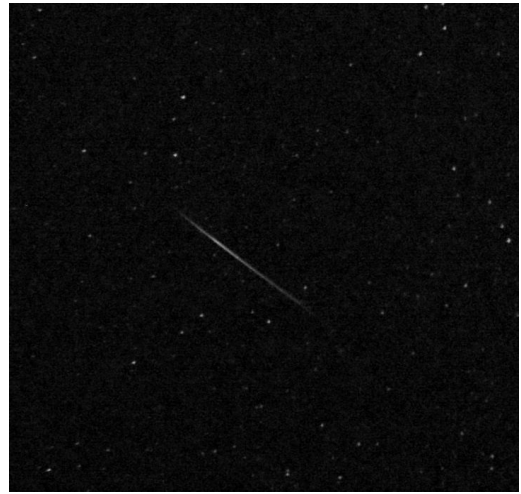

(c)

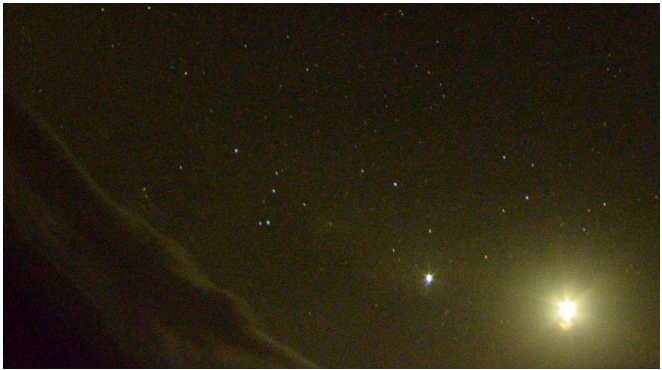

(d)

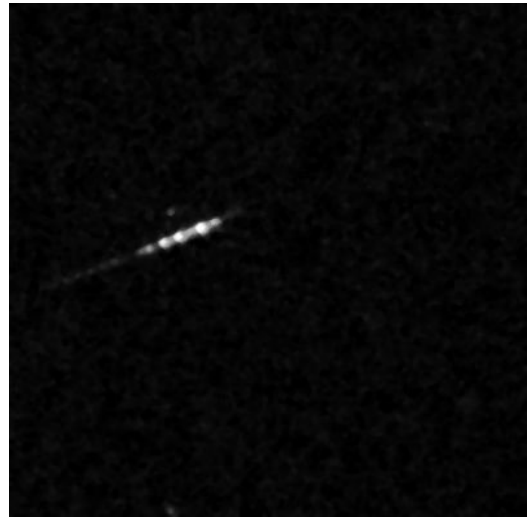

Figure 4

<fig todo, fig 15 in jean>

    Figure 5

(a)

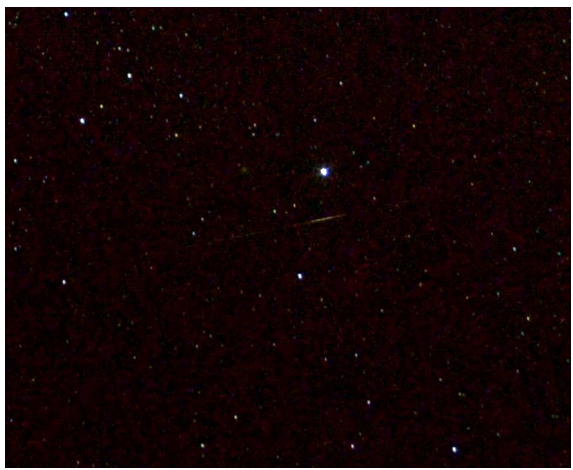

(b)

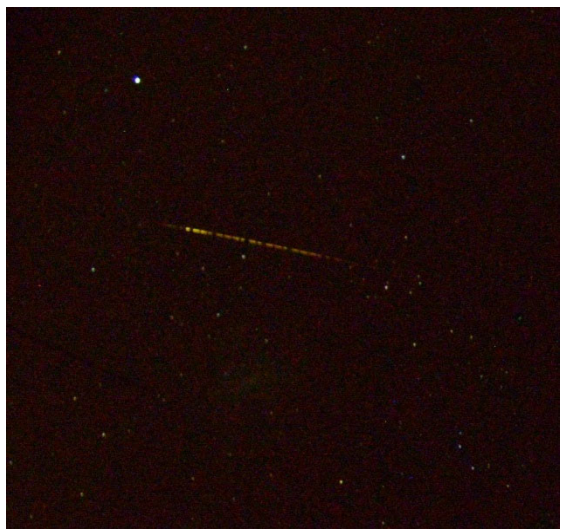

(c)

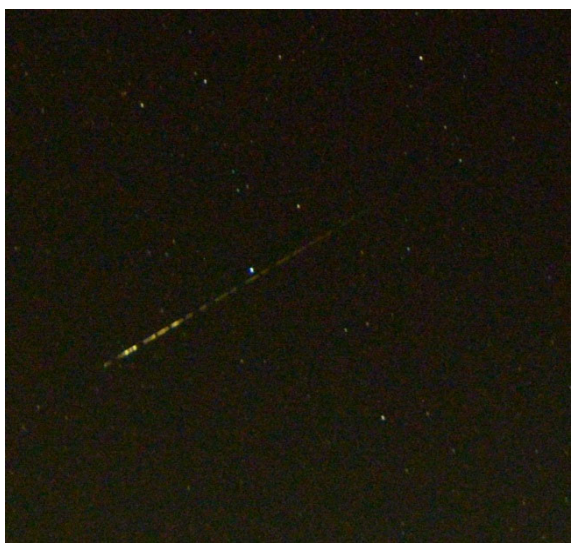

(d)

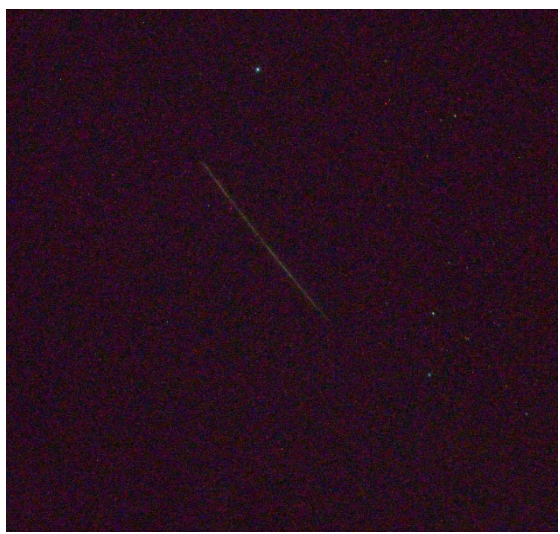

(e)

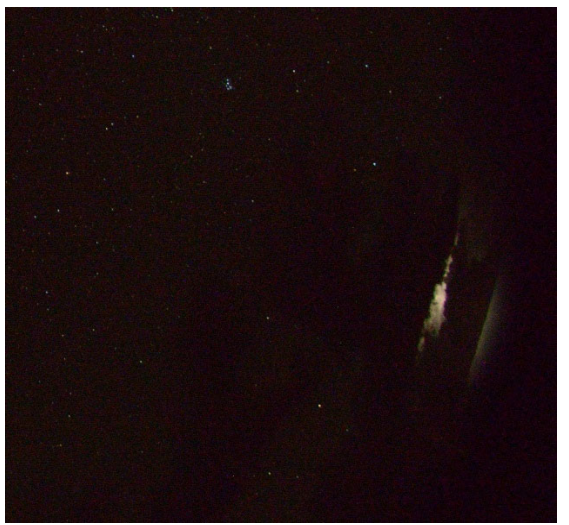

31    Figure 6

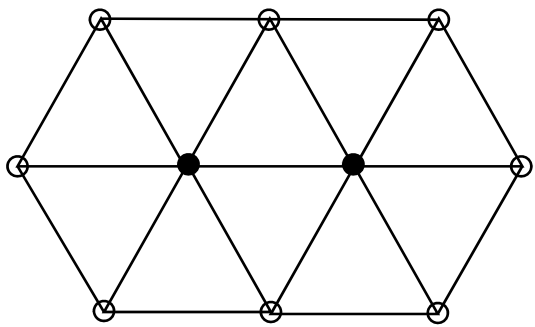

32